\documentstyle[12pt,aaspp4]{article}
\begin{document}
\title{Cepheid and Tip of the Red Giant Branch Distances To the}
\title{Dwarf Irregular Galaxy IC~10\footnote{Based 
on observations made at Palomar Observatory as part of a
continuing collaborative agreement between the California Institute of
Technology and the Jet Propulsion Laboratory.}}

\medskip
\author{\sc Shoko Sakai}
\affil{Kitt Peak National Observatory}
\affil{P.O. Box 26732, Tucson, AZ 85726 \hskip 0.5cm
shoko@noao.edu}
\author{\sc Barry F. Madore}
\affil{NED, California Institute of Technology}
\affil{MS 100--22, Pasadena, CA 91125 \hskip 0.5cm
barry@ipac.caltech.edu}
\author{\sc Wendy L. Freedman}
\affil{The Observatories, Carnegie Institution of Astronomy}
\affil{813 Santa Barbara St. Pasadena, CA 91101 \hskip 0.5cm wendy@ociw.edu}

\bigskip
\centerline{Running Headline: {\it TRGB Distance to IC~10}}

\bigskip
\bigskip
\bigskip
\centerline{Accepted for Publication in {\it Astrophysical Journal}}

\vfill\eject

\begin{abstract}

We present color-magnitude diagrams and luminosity functions of stars
in the nearby galaxy IC 10, based on VI CCD photometry acquired with
the COSMIC prime-focus camera on the Palomar 5m telescope.  The
apparent I-band luminosity function of stars in the halo of IC 10
shows an identifiable rise at I$\approx$ 21.7 mag. This is interpreted
as being the tip of the red giant branch (TRGB) at $M_V \approx$ -4
mag.  Since IC 10 is at a very low Galactic latitude, its foreground
extinction is expected to be high and the uncertainty associated with
that correction is the largest contributor to the error associated
with its distance determination.  Multi-wavelength observations of
Cepheid variable stars in IC 10 give a Population I distance modulus
of 24.1 $\pm$ 0.2 mag, which corresponds to a linear distance of 660
$\pm$ 66 kpc for a total line-of-sight reddening of E(B-V) = 1.16
$\pm$ 0.08 mag, derived self-consistently from the Cepheid data alone.
Applying this Population I reddening to the Population II halo stars
gives a TRGB distance modulus of 23.5 $\pm$ 0.2 mag, corresponding to
500 $\pm$ 50 kpc.  We consider this to be a lower limit on the TRGB
distance.  Reconciling the Cepheid and TRGB distances would require
that the reddening to the halo is $\Delta$E(B-V) = 0.31 mag lower than
that into the main body of the galaxy.  This then suggests that the
Galactic extinction in the direction of IC~10 is $E(B-V) \simeq 0.85$.

\end{abstract}

\it Subject headings: \rm galaxies: individual (IC~10)   --
galaxies: dwarf galaxies -- galaxies: distances
\vfill\eject
\bigskip
\bigskip

\section{Introduction}
IC~10 is a dwarf galaxy located at $\alpha = 00^h 20^m.4$ and 
$\delta = 59^d 18^m$ (2000).
As the nearest example of a post--burst dwarf
galaxy, IC~10 has been recognized as an important object particularly
in studies of the interstellar medium and star formation in
dwarf irregular galaxies.  This galaxy 
has a heliocentric velocity of $-344 \pm 3$ km s$^{-1}$ (RC3, 1991).
However, the distance to IC~10 has been very
poorly determined until recently.
For almost three decades, the distance estimates for IC~10 have
ranged between 1 and 3 Mpc.  This uncertainty is largely attributable
to the fact
that IC~10 is located at a very low Galactic latitude, $b = -3\arcdeg$,
and large extinction corrections need to be applied.
One of the earliest distance estimates was reported
by de Vaucouleurs \& Ables (1965); their value of 1.25 Mpc 
was based on the largest ``ring--like'' HII regions.
Judging from the large HI extent of the galaxy, Roberts (1962) also 
placed this galaxy at 1 Mpc.
There were, however, subsequent studies that suggested a significantly
larger distance for IC~10.
For example, Sandage \& Tammann (1974) reported that its distance was
3 Mpc, based on the size of the three largest HII regions.
Using the HII rings again, de Vaucouleurs (1978) then suggested 
that IC~10 was at 2 Mpc; while an
upper limit of 2.2 Mpc was suggested by Jacoby \& Lesser (1981)
from the observations of a single planetary nebula.
Yahil et al. (1977) concluded that based on this galaxy's degree of 
resolution into stars, it should be located at around 1.5 Mpc.
Bottinelli et al. (1984) used a Tully--Fisher relation to determine
the distance of 2 Mpc.  However, unfortunately 
this was based on B--band photometry
which required a large extinction correction.

Recent observations now suggest that IC~10 is a member of the 
Local Group.
Studies of Wolf-Rayet stars by Massey \& Armandroff (1995) first indicated
that IC10 lies at a distance of only 950 kpc.
Subsequently, 
Saha et al. (1996) discovered Cepheid variable stars,
determining a distance of 830 kpc to IC~10.  Infrared
observations of these same Cepheids by Wilson et al. (1996) reported 
the distance of 820 kpc.
Unfortunately, the color--magnitude diagram of Saha et al. (1996) did 
not penetrate deep enough to probe the red giant branch stars, even
though they did in fact visually detect the background ``Baade's sheet''
of red stars.  These red giant branch stars can provide
an independent, Population II measure of the distance to IC~10;
and that is the subject of this paper.
We also present the $V$ and $I$ Cepheid data from our frames
and the distance using
their period--luminosity relation.
Furthermore, we derive the reddening correction estimate from the 
multi--wavelength observations of the Cepheid variables, compiled from
new and previously published data.

As part of a continuing effort to obtain consistent distances to all
the Local Group galaxies, we present in this paper, the detection and
measurement of the tip of the red giant branch (TRGB) in IC~10.
The TRGB marks the helium core flash, which is detected in the $I-$band
luminosity function as an abrupt discontinuity.
The TRGB has been demonstrated observationally
and shown theoretically to be an excellent distance indicator that is 
as accurate as the period--luminosity relation of Cepheid variable stars
(Frogel, Cohen \& Persson 1983, DaCosta \& Armandroff 1990, 
Lee, Freedman \& Madore 1993, Madore, Freedman \& Sakai 1997 and references
therein).  The major advantage of the TRGB method over Cepheid variables
is that its application is much quicker.  In principle, only one epoch
of observations is needed.  The method can also be applied to
any morphological type of galaxies.
However, the limitation is that an independent estimate of reddening is
required.
This could pose additional uncertainties, 
especially in this particular application of IC~10, in which the
errors in the reddening correction dominate the error in its 
distance estimate due to its low Galactic--latitude location.

\section{Observations}

Observations of IC~10 were made at Palomar Observatory using the
Hale 5m telescope on two consecutive nights, October 5th and 6th,
1996. All the observations were done under photometric conditions,
with moderate seeing ($\sim1.2$ arcsec).
The Carnegie Observatories Spectroscopic Multislit and Imaging Camera
(COSMIC, Kells et al. 1998), 
a prime focus camera, was used to obtain $V$ and $I-$band images, 
with total exposure times being
480 and 600 sec for the first night, and 720 and 1080 sec for
the second night.  The data were debiased and 
flatfielded using standard
reduction methods.  Stellar photometry was obtained using the point--spread
function fitting packages DAOPHOT and ALLSTAR (Stetson 1987), which 
use automatic star finding algorithms.  A point spread function,
as determined from bright, isolated stars in the same field, was then fit
to extract total magnitudes.  

A set of $V$ and $I$ standard stars, selected from Landolt's catalog (1992), 
were observed at least once every hour throughout both nights, and
the two nights were calibrated independently.  
The photometry comparison indicates that the 
zero--point calibrations of two nights are in excellent agreement; 
for both $V$ and $I$, the magnitudes of brightest stars agree 
to within 0.01 mag.
In the following sections, however, we will only present the data from
the second night of observations, mainly because the telescope had
moved slightly during the exposures of the first night.  Thus, rather
than combining the data from two nights, our analysis will focus on the
second-night data only.

\section{Luminosity Function and Color Magnitude Diagram}

Figure 1 shows the COSMIC $I-$band image of IC10.  
We refer to the main body region within the inner ellipse as Region 1.
The annular region between inner and outer ellipses is called Region 2,
while the remainder of the frame is referred to as Region 3.
In Figures 2a-c, a $(V-I)$ vs. $I$ color-magnitude diagrams (CMD) of three
regions in IC10 are shown.

In the main body, as observed from the CMD in Figure 2a, 
a red giant branch is present, as well as a sparse and ill--defined
blue main sequence stellar population around $V-I \simeq 1.0$ mag extending
from $I = 20.5$ down to 22.5 mag.  Also present are some intermediate-age 
asymptotic giant branch (AGB) stars in the region slightly brighter than 
the RGB.  The red giant branch is more clearly demonstrated in 
Figure 2b, for the halo region of IC~10.  However, AGB stars are also present
in this CMD and care must be exercised to discriminate between the position 
of the TRGB and AGB in the luminosity function (below).
In Regions 2 \& 3, the blue
main sequence stars are no longer present, especially in Region 3; however,
the foreground stars start dominating the CMD region at $V-I \simeq 1-3$
mag at brighter magnitudes between $I = 17$ and $20.5$ mag.

Histograms in Figure~3 show the $I-$band luminosity functions for the 
stellar populations found in Region 1 (left) and Regions 2 \& 3 of IC~10.
In the Region 1 luminosity function, we see no distinct discontinuity at any
point.  In contrast, 
the main characteristic of the halo $I-$band luminosity function (Regions
2 and 3) is the jump by nearly 50\% in counts (between adjacent bins 0.1 mag
in width), at $I \simeq 21.7$ mag.
This, we believe, marks the tip of the red giant branch, which is
the focus of Section 5.
There is also a jump, though significantly smaller than the one at 21.7 mag,
at $I \sim 21.4$ mag.  We ascribe this feature to the AGB population intrinsic
to IC~10.

In the next section, we discuss the distance to IC~10 derived from
Cepheid variable stars, which is then compared with the tip of the
red giant branch method in the subsequent section.

\section{Cepheid Variable Stars in IC~10}

Saha et al. (1996) discovered 13 variable star candidates in IC~10,
nine of which were identified as Cepheids or Cepheid-like stars.
However, their observations were undertaken using Gunn $gri$ filters.
Here, we have recovered some of these Cepheids on our COSMIC
frames and their $V$ and $I$ magnitudes are presented in Table 1.
Unfortunately, the brighter Cepheid variables were saturated
on our images, so we were unable to photometer some of the
candidates.  The $V$ and $I$ Cepheid data are plotted in Figure 4.
They represent random--phase period-luminosity ($PL$)
relations; no phase corrections or averaging were applied.

The absolute calibrations for the $PL$ relations are adopted from 
Madore \& Freedman (1991) which are based on a consistent set of
25 Large Magellanic
Cloud (LMC) Cepheid variables with $BVRGIJHK$ observations and expressed as:

\begin{equation}
M_V = -2.88 (\pm 0.20) (\log P - 1.00) - 4.11 (\pm 0.09) [\pm 0.29],
\end{equation}

\begin{equation}
M_I = -3.14 (\pm 0.17) (\log P - 1.00) - 4.84 (\pm 0.06) [\pm 0.21].
\end{equation}

These calibrations assume values of $(m-M)_0 = 18.50 \pm 0.10$ mag
and $E(B-V) = 0.10$ mag for the distance modulus and reddening of the LMC.
The apparent distance modulus for the IC~10 data at each wavelength was 
determined by
minimizing the {\it rms} deviations of the observed data about the
ridge line, with the slopes fixed to those given by the above equations.
For $V$ and $I$, we obtain distance moduli, respectively, of
$(m-M)_v = 27.87 \pm 0.11$ and $(m-M)_i = 25.98 \pm 0.14$; the shortest--period
Cepheid (V6) with $P = 8$d was omitted from these calculations, given its anomalous color and
also as to avoid the possible
influence of overtone pulsators.

Near--infrared magnitudes of four Cepheid variables in IC~10 are listed in
Table 2 of Wilson et al. (1996).  We follow the same procedures as
above to obtain apparent distance moduli in $JHK$, using the following
absolute calibration, again provided by Madore \& Freedman (1991) based
on 25 LMC Cepheids:

\begin{equation}
M_J = -3.31 (\pm 0.11) (\log P - 1.00) - 5.29 (\pm 0.05) [\pm 0.16],
\end{equation}

\begin{equation}
M_H = -3.37 (\pm 0.10) (\log P - 1.00) - 5.65 (\pm 0.04) [\pm 0.14],
\end{equation}

\begin{equation}
M_K = -3.42 (\pm 0.09) (\log P - 1.00) - 5.70 (\pm 0.04) [\pm 0.13].
\end{equation}

The $J$, $H$ and $K$ apparent distance moduli are $(m-M)_J = 25.62 \pm 0.23$,
$(m-M)_H = 25.05 \pm 0.29$ and $(m-M)_K = 24.39 \pm 0.34$ respectively.

Following the procedure outlined in detail by Madore \& Freedman (1991),
a reddening law, consistent with Cardelli, Clayton \& Mathis (1989), was 
fitted to the
$VIJHK$ multi--wavelength apparent distance moduli, as shown in Figure 5.
Extrapolating to $\lambda^{-1} = 0$, we obtain a true distance modulus
of $24.10 \pm 0.19$ mag (660 $\pm$ 63 kpc), with a reddening of
$E(B-V) = 1.16 \pm 0.08$ mag.  
Since the determination of the true distance modulus requires finding the
minimum $\chi^2$ solution in the extinction/modulus plane, the
errors in these two variables are dependent on each other.  Thus the
uncertainties in the distance modulus are illustrated in the enclosed box
in Figure 5 as $\chi^2$ contour ellipses, ranging from 1 to $3-\sigma$.

\section{Detection of the Tip of the Red Giant Branch}

The TRGB marks the core helium flash of old, low--mass stars.
These stars evolve up the red giant branch, but almost instantaneously 
change their physical characteristics upon ignition of helium, 
which in turn appears as a 
sudden discontinuity in the luminosity function.
In the $I-$band ($\sim 8200$\AA), the tip is observed at $M_I \simeq -4$ mag,
and this magnitude has been shown both observationally and theoretically
to be extremely stable; it varies only by $\sim$0.1 mag for ages
2 -- 15 Gyr, and for metallicities between $-2.2 <$ [Fe/H] $< -0.7$ dex,
(the range spanned by the Galactic globular clusters).
Here, we use the calibration determined by Lee et al. (1993) which is
based on the observations of four Galactic globular clusters by
Da Costa \& Armandroff (1990).

The foreground extinction value for IC~10 has been a major obstacle when 
determining the distance to this galaxy accurately, as IC~10 is located 
at the very low Galactic latitude of only $b=-3\fdg3$.
The estimate given by de Vaucouleurs \& Ables (1965) of $E(B-V) = 0.87$
mag was used as a standard value for many years.
Other estimates ranged from $E(B-V) = 0.4$ mag (de Vaucouleurs 1978) 
up to $1.7 - 2.0$ mag (Yang \& Skillman 1993).
More recent studies by Massey \& Armandroff (1995) used 
the Wolf--Rayet stars and
the location of the main sequence blue plume to determine the
foreground extinction value, and concluded $E(B-V) = 0.75 - 0.80$ mag.
In this paper, we derive a value of $E(B-V) = 1.16 \pm 0.08$ mag, which
was obtained from the multi--wavelength Cepheid observations in the 
previous section.
Using conversions of $A_V / E(V-I) = 2.45$ and $R_V = A_V/E(B-V) = 3.2$ 
(Dean, Warren \& Cousins (1978), Cardelli et al. (1989) and Stanek (1996)),
we obtain $A_V = 3.71 \pm 0.26$ and $A_I = 2.19 \pm 0.15$.  
The uncertainty in the reddening
estimate is one of the largest sources of systematic errors in 
the IC~10 distance.  We note, however, that the extinction for the red giant
branch stars is likely to be smaller than $E(B-V) = 1.16$ mag. 
The Cepheid variables
are usually found in and around the star--forming regions of the main
body of the galaxy, which probably suffer from more reddening than 
the halo region where the RGB stars are observed.  
However, in the case of IC10, the foreground reddening dominates that internal
to the galaxy.

The top panels in Figure~6 show $I-$band luminosity functions of the
stars in Region 1 (left) and Regions 2 and 3 which were shown in Figure~3, 
but smoothed by a variable Gaussian whose dispersion 
is the photometric error for each star detected.  
A Sobel edge--detection filter is applied to the smoothed luminosity
functions following an equation:
$E(m) = \Phi(I + \sigma_m) - \Phi(I - \sigma_m)$, where $\Phi(m)$ is
the luminosity function at magnitude defined at $m$.  
For the details of the Sobel filter application, readers
are referred to the Appendix of Sakai, Madore \& Freedman (1996).
The filtered function output are shown in the bottom panel of Figure~6.
The position of the TRGB is indicated by the highest peak
in the filter output.  
The data used in Figure 6 include all the stars found in specified regions.
Here, however, we are interested in the red giant branch luminosity function.
Using the $(V-I)$ color information, we select a subsample of stars with
$2.5 \leq 
V-I \leq 3.0$, effectively excluding the bluer foreground stars which 
are merely noise in our TRGB application.  
The results are shown in Figure 7a where both the histograms and smoothed
luminosity functions are used to illustrate the position of the tip.
The position of the TRGB is indicated by both the significant jump
in the number counts, and also by the prominent peak in the edge--detection
filter output.
To demonstrate the effectiveness of this scheme in which we selectively
use only a subset of the RGB population, 
we show the luminosity function histograms and filter
output for RGBs of redder and bluer regions in Figure 7b.
For the bluest sample ($2.0 \leq V - I \leq 2.5$), although the TRGB can be
still detected at $I \simeq 21.7$ in the filter output, the luminosity
function histogram does not exhibit any significant corresponding
discontinuity.  
For the redder sample of $3.0 \leq V - I \leq 3.5$, it is nearly
impossible to visually identify the TRGB position in its luminosity function.
Using the $2.5 \leq (V-I) \leq 3.0$ sample, we conclude that
the TRGB is detected at $I = 21.70 \pm 0.15$ mag.
The `full width half maximum'' of the output response peak profile is used to 
estimate an uncertainty of $\pm 0.15$ mag on the apparent modulus.

\subsection{TRGB Distance to IC~10}

To calculate the true modulus to IC~10, 
we use the TRGB calibration of Lee et al. (1993), according to which
the tip distance is determined via the relation $(m-M)_I = I_{TRGB} -
M_{bol} + BC_I$, where both the bolometric magnitude ($M_{bol}$) and
the bolometric correction ($BC_I$) are dependent on the color of
the TRGB stars.  They are defined by: $M_{bol} = -0.19[Fe/H] - 3.81$ and
$BC_I = 0.881 - 0.243(V-I)_{TRGB}$.  The metallicity is in turn expressed
as a function of the $V-I$ color: $[Fe/H] = -12.65 + 12.6(V-I) - 3.3(V-I)^2$.
The colors of the red giant stars, corrected for reddening, range from
$(V-I)_0 = 0.6 - 1.6$ (see Figure~2), which gives
the TRGB magnitude of $M_I = -4.00 \pm 0.10$.
We thus derive the TRGB distance modulus of IC~10 to be
$(m-M) = 23.51 \pm 0.19$ mag, {\it adopting the reddening of $E(B-V) = 1.16$,
derived from the combined observations of the optical and IR Cepheid variable 
stars}.
This corresponds to a linear distance of $500 \pm 48$ Kpc.
The sources of errors include the uncertainties in (1) the tip position
(0.05 mag), (2) reddening (0.15 mag) and (3) TRGB calibration (0.10 mag).
This is the lower limit on the TRGB distance as the
extinction in the halo is likely less than that in the main body of the galaxy
where the Cepheid variables are detected.


The fact that the TRGB method requires an {\it independent} estimate
of the reddening is a clear disadvantage,
unlike the multi--wavelength Cepheid observations. This is especially
problematic in the case of IC~10.
In Table 2 we present various estimates of E(B-V) for IC 10 and the
corresponding values for $A_V$ and $A_I$.  Also tabulated are the true
modulus and linear distance one would obtain using our TRGB magnitude
combined with the suggested reddening. Distance estimates cover a
factor of 4$\times$, ranging from 230 up to 950 kpc, depending on the
adopted reddening.  The Cepheid-based distances derived by Saha et al.
(1996: 830 $\pm$ 120 kpc) and Wilson et al. (1996: 820 $\pm$ 80 kpc)
reduce to 660 $\pm$ 63 kpc when JHK data of Wilson et al.  are
combined with the VI data, reported here.  This is directly a result
of an increased reddening estimate derived from the multiwavelength
data.  

We consider the TRGB distance of 500 kpc that we obtain using the Cepheid
reddening to be a lower limit, given that the line-of-sight
reddening appropriate to the halo of IC 10 (where the red giant stars
used in our analysis are located) is expected to be smaller than that
of the main body of the galaxy where the Population I Cepheids, dust
and gas are primarily concentrated.  
Without an independent measure of the reddening along the
line-of-sight to the halo of IC~10, one alternative is to
adopt the Cepheid distance, and then deduce the line--of--sight
reddening to the halo.
These two alternatives are illustrated by the
color magnitude diagrams in Figure 8.
On the left--hand side is the CMD in which the $V$ and $I$ magnitudes
have been shifted by the TRGB distance derived using the extinction derived
in this paper from the optical/IR Cepheid observations.
Overplotted lines represent red giant branches of six Galactic
globular clusters presented in Da Costa \& Armandroff (1990);
they do not quite match the IC~10 RGB.
It is clear that the adopted distance modulus of 23.51 does
not yield a consistent view.
On the other hand, we can adopt the Population I Cepheid distance
modulus of $m-M = 24.10$ as the true distance to IC~10.
This would then mean the line-of-sight reddening to the
halo of IC~10 becomes $E(B-V)_{\mbox{foreground}} \sim 0.85$~mag.
The corresponding CMD is shown on the left--hand side of Figure 8.
The IC~10 RGB matches well with those of the Galactic globular clusters,
suggesting that the lower extinction in the halo region is 
a more sensible value to adopt here.

Even for the case presented on the right in Figure 8, we note that
the Galactic globular cluster isochrones do not match the IC~10
RGB population very well. 
For example, there are a few stars observed at metallicities
higher than the most metal--rich isochrone, and also in the 
bluer region.  
The metallicity of IC~10, measured from the observations of HII regions,
is reported as $12 + \log (O/H) = 8.2$, which is similar to NGC~6822.
Although this is a Pop I metallicity measurement, we do not expect
the Pop II RGB metallicity to be much higher than this.
The detection of stars ``outside'' the globular cluster isochrone range
is likely due to a combination of several factors.
First, the photometric errors for stars of fainter magnitudes at $I \geq 22.0$  
reach $0.3 - 0.5$ mag.  Thus, the stars with unreasonable colors could be
simply due to uncertain photometric results.
Second, the crowding probably plays a major role.
When examining those stars in question more closely on the CCD images,
a significant number of them in fact are located close to the foreground
stars, star clusters, or HII regions.
Variable reddening undoubtedly affects the photometric results as well.
From our data alone, it is impossible to correct for such an effect,
not to mention to estimate the degree of the reddening variation.


\section{Summary}

Using $V$ and $I$ photometry, the distance to a dwarf irregular
galaxy, IC~10, has been determined using both the multi--wavelength 
Cepheid $PL$ relation (Pop I) and the tip of the red giant branch method
(Pop II). 
Adopting a total line--of--sight color excess of
$E(B-V) = 1.16 \pm 0.19$ mag based on the Cepheid photometry, we derive
the Population I distance of
$(m-M) = 24.10 \pm 0.19$ (660 $\pm$ 63 kpc).
Adopting this reddening yields the Population II distance
of $(m-M) = 23.41 \pm 0.19$ ($481 \pm 45$ kpc), which
is a lower limit as the line--of--sight extinction
in the halo region is smaller than that in the main body of the galaxy.
If we adopt the Cepheid distance as the true distance to IC~10,
it would then imply that the foreground reddening in the 
line--of--sight to the halo of IC~10 is $E(B-V)_{\mbox{foreground}} 
\sim 0.85$ mag.

This work was funded by NASA LTSA program, NAS7-1260, to SS.
BFM was supported in part by the NASA/IPAC Extragalactic Database.

\newpage

{\bf Figure Captions}

Figure 1: An $I-$band image of IC~10.  
Three regions used in the analysis are separated by the ellipses drawn.

Figure 2: An $I - (V-I)$ color magnitude diagram for Region 1 (a), Region
2 (b) and Region 3 (c).

Figure 3: Histograms showing $I-$band luminosity functions for the 
main body of the galaxy, Region 1 {\it (top)} and for the
halo region {\it (bottom)}.

Figure 4: $V$ and $I$ period--luminosity relations for Cepheid variable stars
detected on the COSMIC images.

Figure 5: Multiwavelength fit of a Galactic reddening law to $VIJHK$ apparent
distance moduli for IC~10.  We obtain a true distance modulus of 24.10 mag.
The inset box contains a contour plot showing the $\chi^2$ values from
fits to determine the true distance modulus.  

Figure 6: Smoothed $I-$band luminosity functions {\it (top)}, and the
edge--detection filter response function {\it (bottom)}.  The position of the
TRGB is indicated by the highest peak in the response function.
The three contour levels represent 1, 2 and 3$\sigma$ error ellipses.

Figure 7: $I-$band luminosity functions and the filter
response functions for red giant branch stars only.

Figure 8: Color--magitude diagrams of IC~10 Region 3, shifted by the
distance modulus and reddening as indicated on top of each plot.

\begin{tabular}{ccccccc}
\multicolumn{7}{c}{\bf Table 1: $VI$ Cepheid data} \\
\cr\hline
\hline\cr
\multicolumn{1}{c}{Cepheid}&
\multicolumn{1}{c}{Period}&
\multicolumn{1}{c}{JD}&
\multicolumn{1}{c}{$V$}&
\multicolumn{1}{c}{$\sigma_V$}&
\multicolumn{1}{c}{$I$}&
\multicolumn{1}{c}{$\sigma_I$}\\
\hline\cr

V1  & 19.12 &  2450361.7 &  22.59  &   0.04 &   19.91 &    0.09  \cr
V1  & 19.12 &  2450362.8 &  22.45  &   0.04 &   19.45 &    0.10   \cr
V2  & 11.87 &  2450362.8 &  23.47  &   0.06 &   22.08 &    0.14   \cr
V2  & 11.87 &  2450362.8 &  23.46  &   0.06 &   21.43 &    0.12   \cr
V4  & 57.60 &  2450361.7 &  21.65  &   0.04 &   19.16 &    0.09   \cr
V4  & 57.60 &  2450362.8 &  21.48  &   0.05 &   18.84 &    0.10   \cr
V5  & 35.29 &  2450361.7 &  21.57  &   0.05 &   19.24 &    0.13   \cr
V5  & 35.29 &  2450362.8 &  21.52  &   0.05 &   18.64 &    0.10   \cr
V6  &  8.09 &  2450362.8 &  25.72  &   0.54 &   19.38 &    0.76   \cr
V9  & 53.36 &  2450361.7 &  22.31  &   0.03 &   18.56 &    0.07   \cr
V9  & 53.36 &  2450362.8 &  22.05  &   0.05 &   18.59 &    0.10   \cr
V11 & 90.70 &  2450361.7 &  21.45  &   0.03 &   18.54 &    0.08   \cr
V11 & 90.70 &  2450362.8 &  21.37  &   0.04 &   18.06 &    0.09   \cr
V12 & 48.22 &  2450361.7 &  22.23  &   0.03 &   19.29 &    0.08   \cr
V12 & 48.22 &  2450362.8 &  22.11  &   0.04 &   19.13 &    0.10   \cr

& & & \\
\hline\cr
\end{tabular}

\begin{deluxetable}{lccccc}
\tablewidth{0pc}
\tablecaption{\bf Table 2: Extinction Estimates for IC~10}
\tablehead{
\colhead{Source}&
\colhead{$E(B-V)$}&
\colhead{$A_V$}&
\colhead{$A_I$}&
\colhead{$\mu_0$}&
\colhead{Distance (kpc)}}

\startdata
Cepheids(VIJHK)\tablenotemark{1}   &  1.16   &  3.71 &    2.19 &   23.51 &  504 \nl
Wolf--Rayet Stars\tablenotemark{2} &  0.75 -- 0.8  &   2.4 -- 2.6   &  1.4 -- 1.5  &  24.3 -- 24.2 &  724 -- 691 \nl
Integrated $B-V$ color\tablenotemark{3}  &   0.87  &   2.78 &   1.64 &   24.06 &  649 \nl
HII rings\tablenotemark{4}  &   0.4  &   1.3  &   0.8 &   24.9 &  955 \nl
HII regions\tablenotemark{5}   &   1.7 -- 2.0  &   5.4 -- 6.4  &  3.2 -- 3.8 &   22.5 -- 21.9 &  316 -- 240 \nl
Cepheids(JHK)\tablenotemark{6} & 0.6 -- 1.1 & 1.6 -- 3.5 & 0.9 -- 2.1 & 24.8 -- 23.6 & 912 -- 525 \nl
Cepheids(gr)\tablenotemark{7} & 0.97 & 3.10 & 1.83 & 23.87 & 594 \nl
\tablerefs{(1) this paper (2) Massey \& Armandroff 1995 (3) de Vaucouleurs \& Ables 1965
(4) de Vaucouleurs 1978 (5) Yang \& Skillman 1993 (6) Wilson et al. 1996 (7) Saha et al. 1996
}
\enddata
\end{deluxetable}

\begin{figure}
\plotone{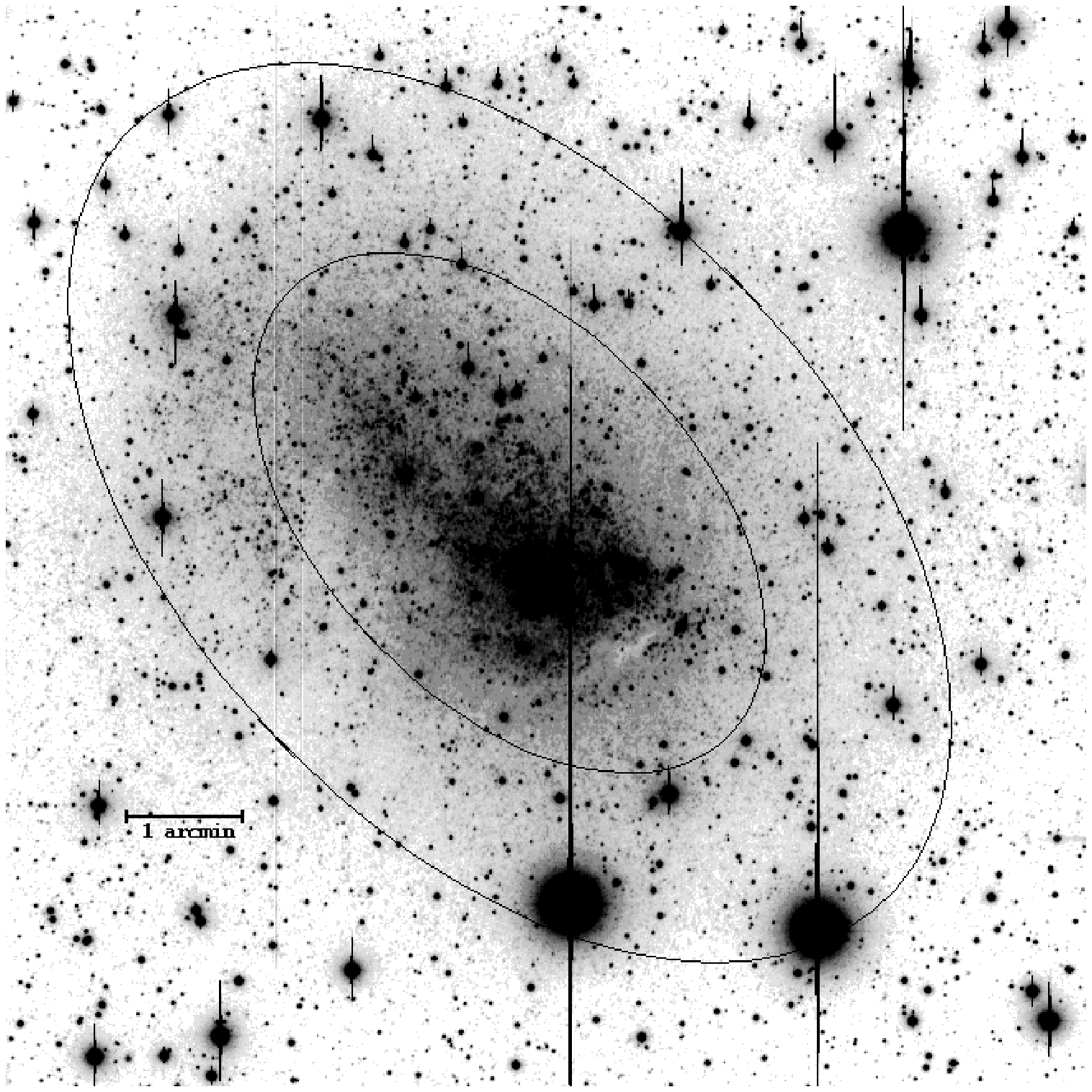}
\end{figure}

\begin{figure}
\plotone{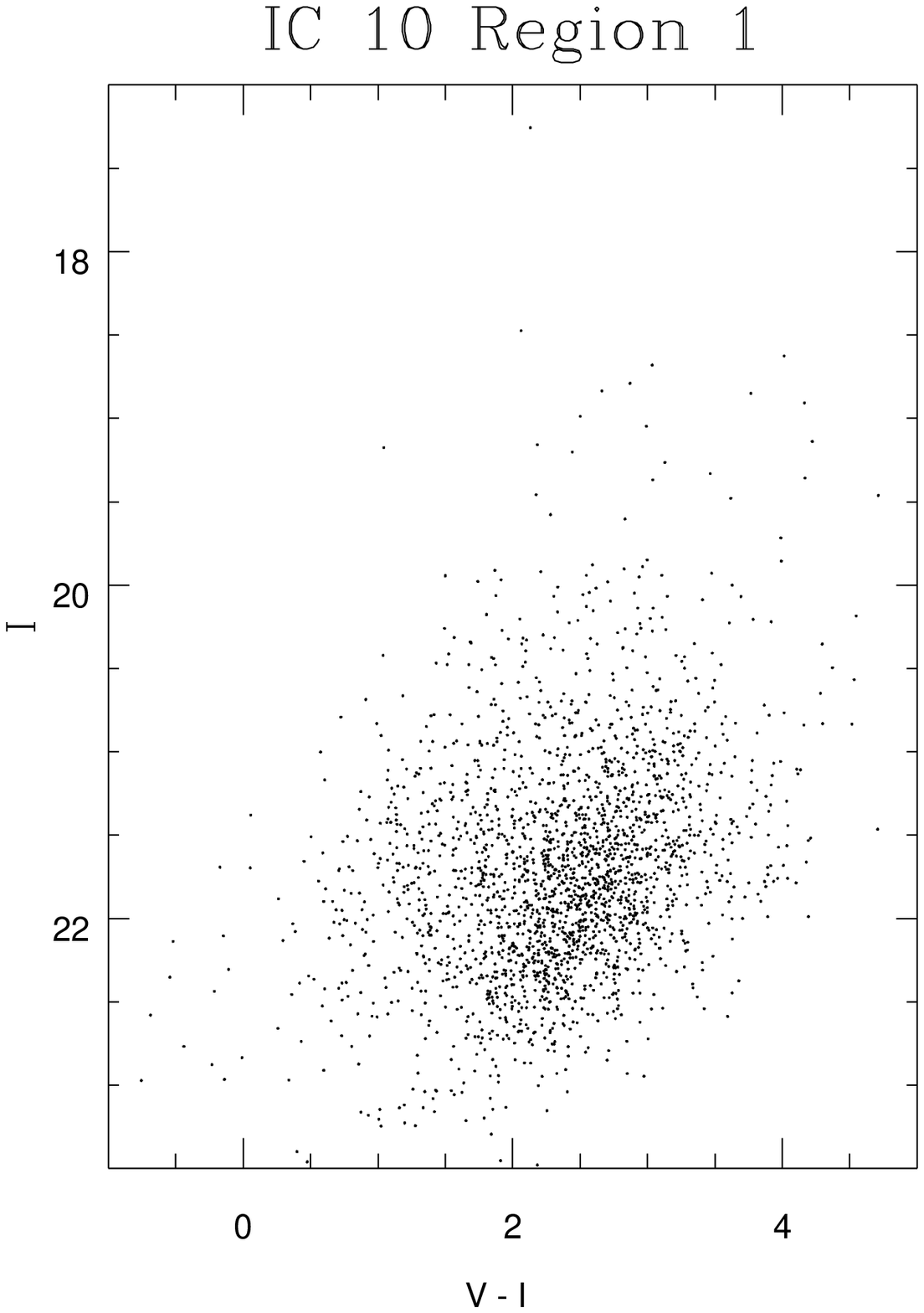}
\end{figure}

\begin{figure}
\plotone{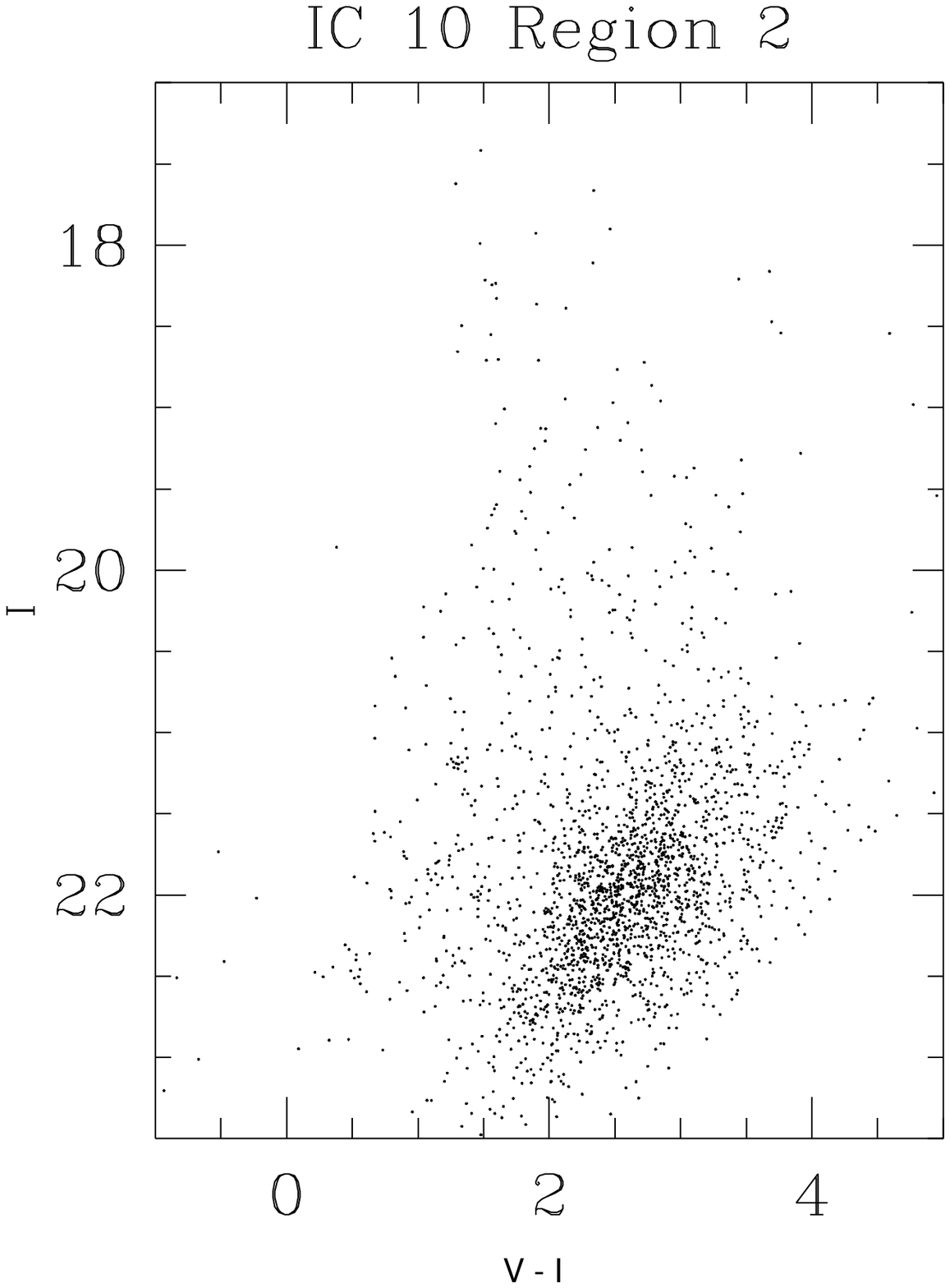}
\end{figure}

\begin{figure}
\plotone{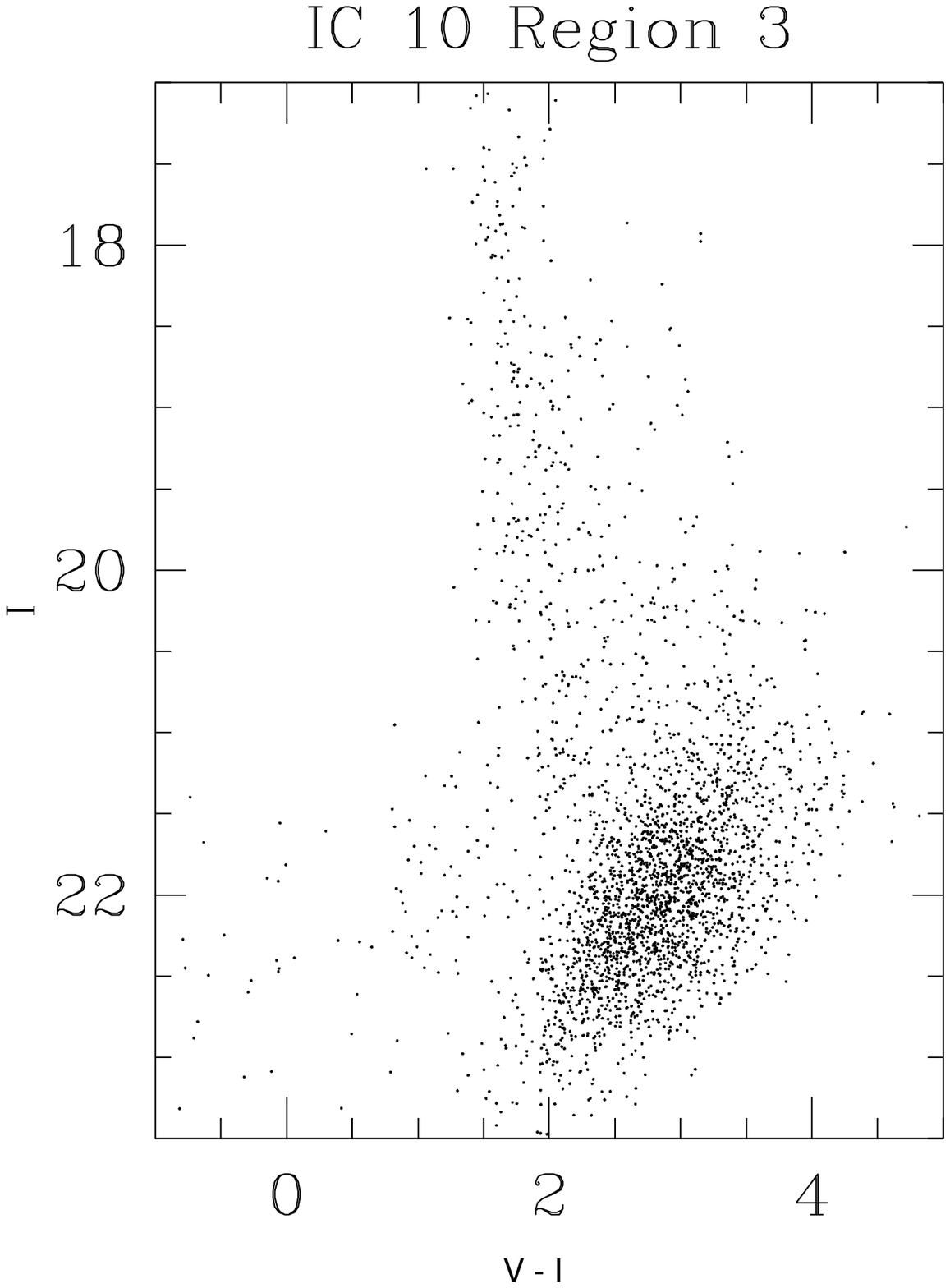}
\end{figure}

\begin{figure}
\plotone{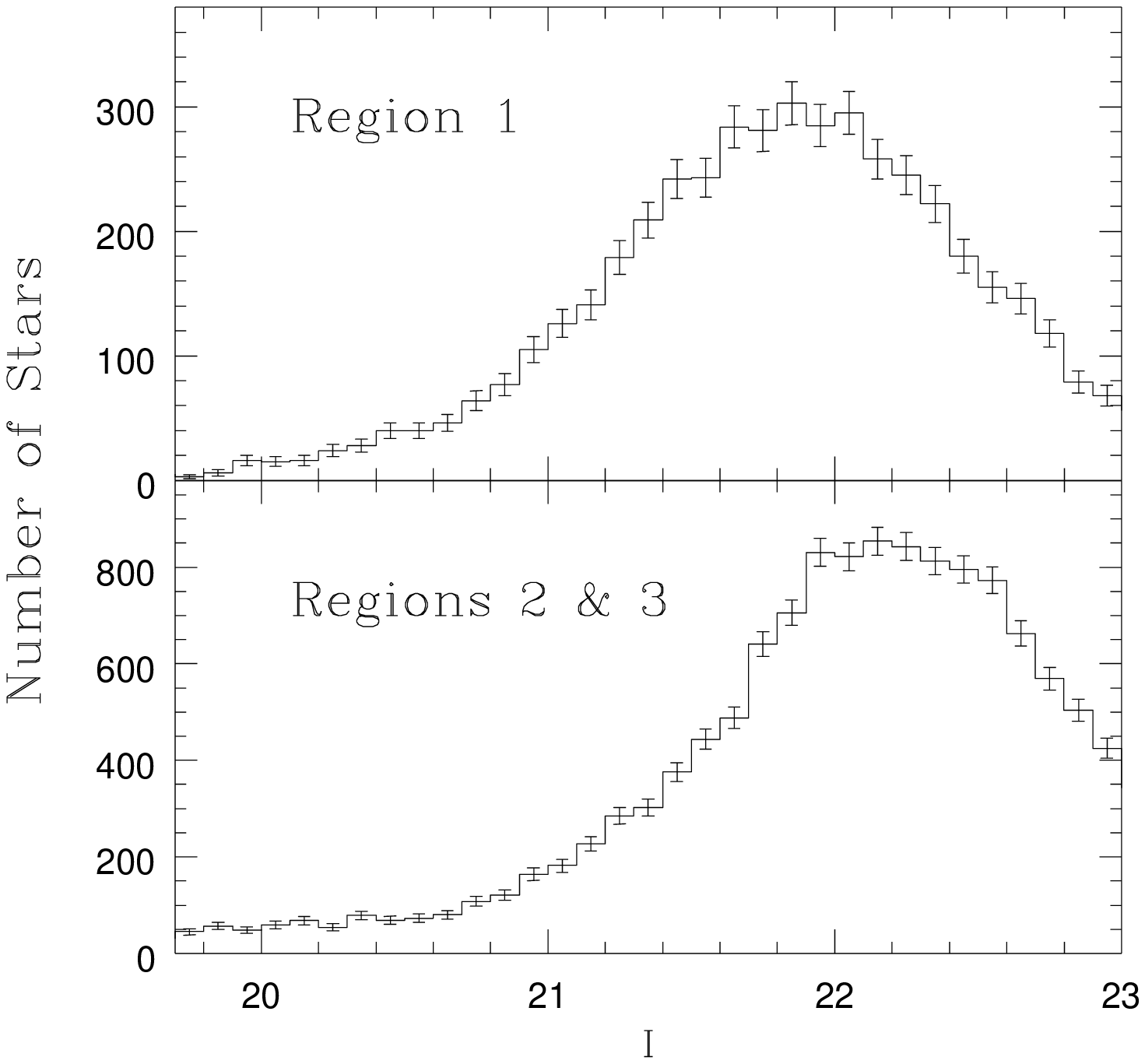}
\end{figure}

\begin{figure}
\plotone{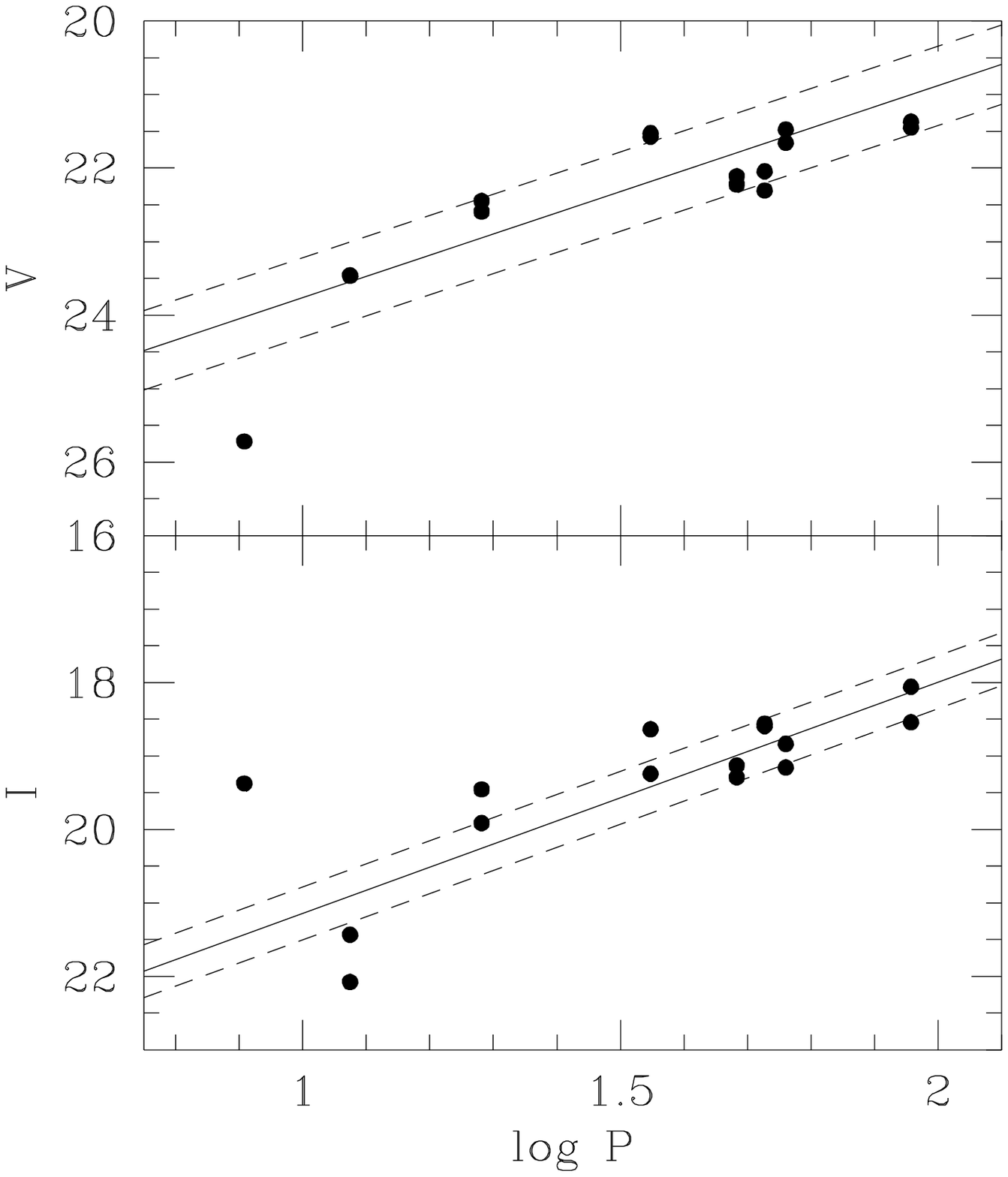}
\end{figure}

\begin{figure}
\plotone{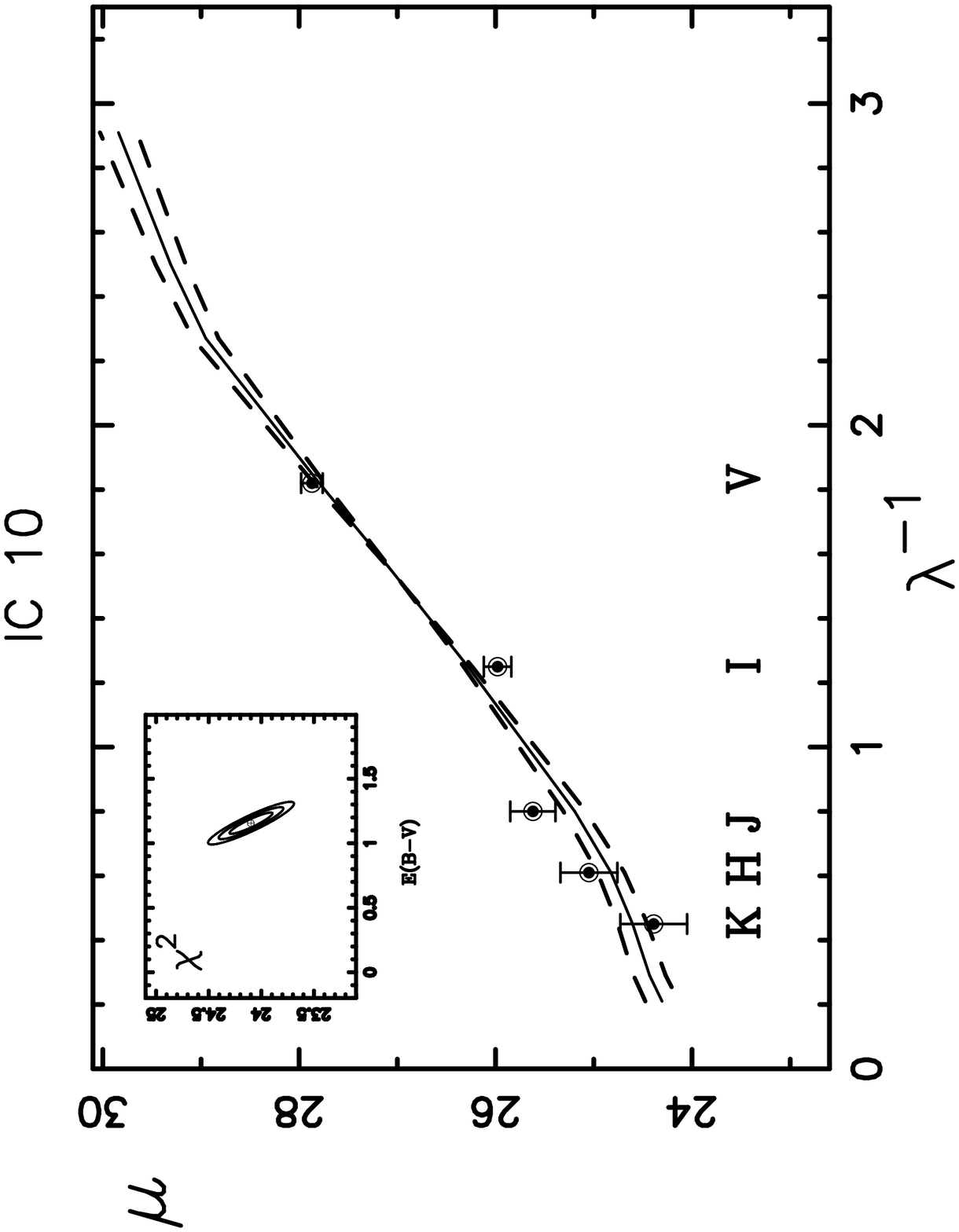}
\end{figure}

\begin{figure}
\plotone{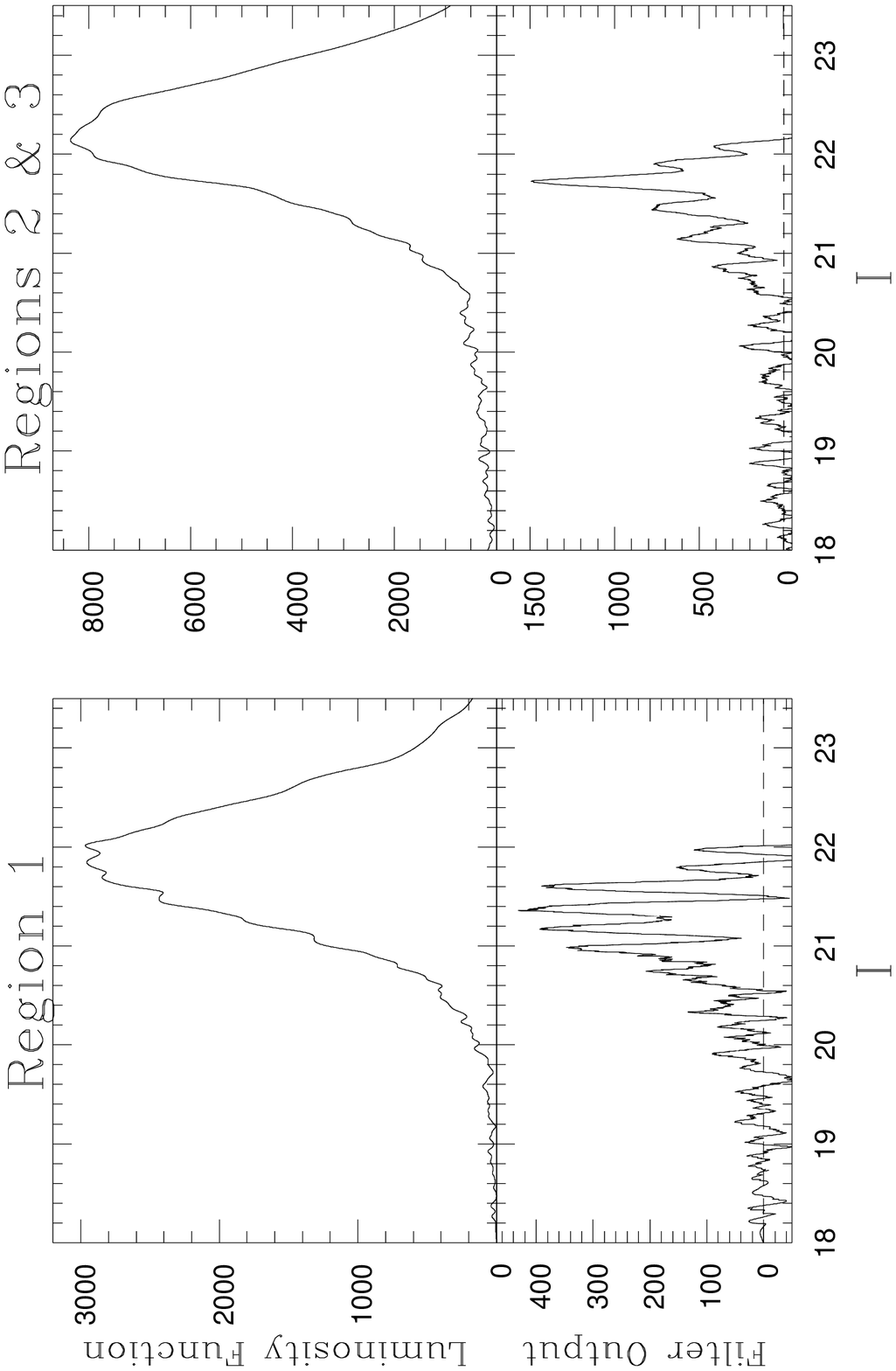}
\end{figure}

\begin{figure}
\plotone{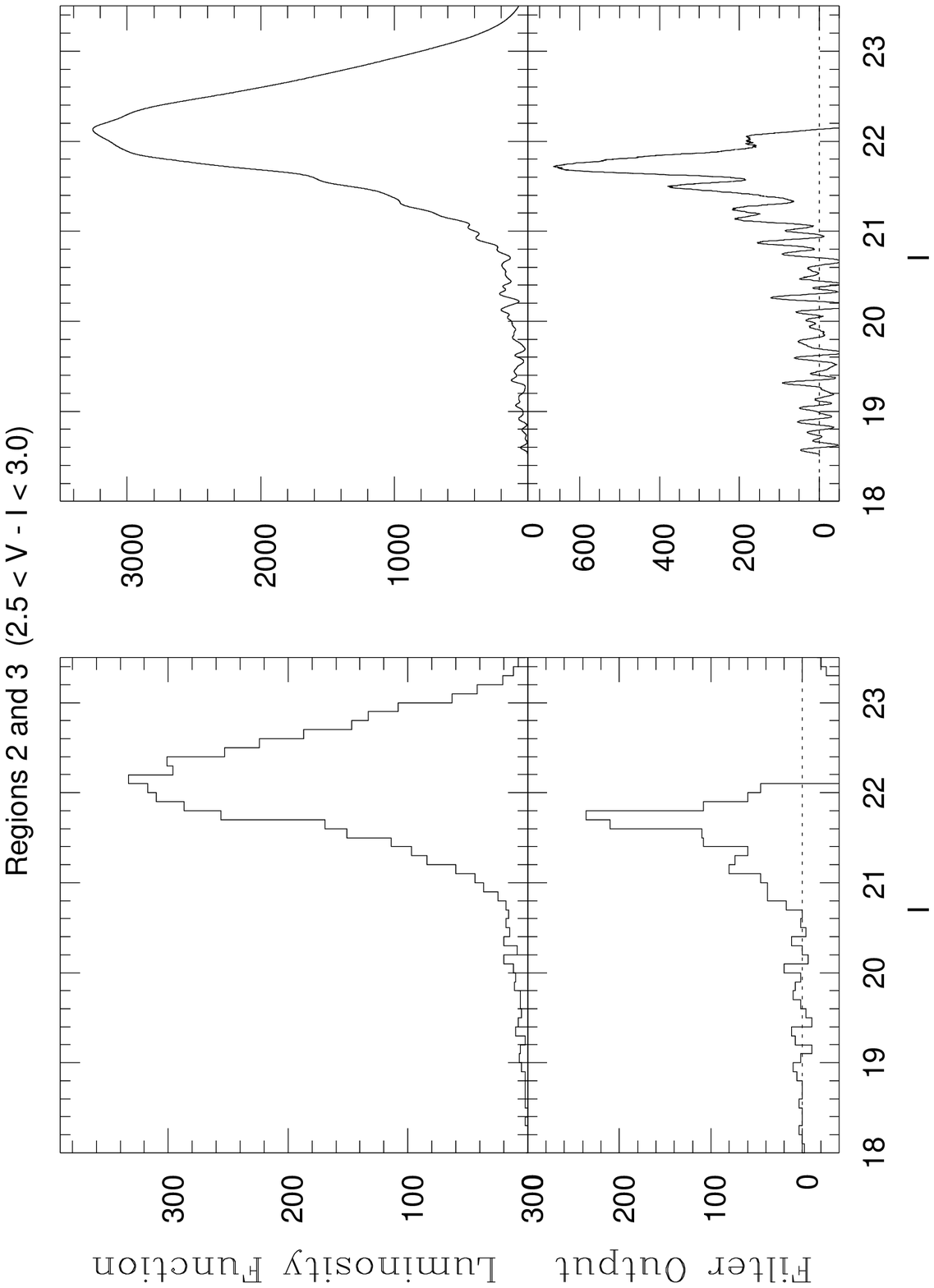}
\end{figure}

\begin{figure}
\plotone{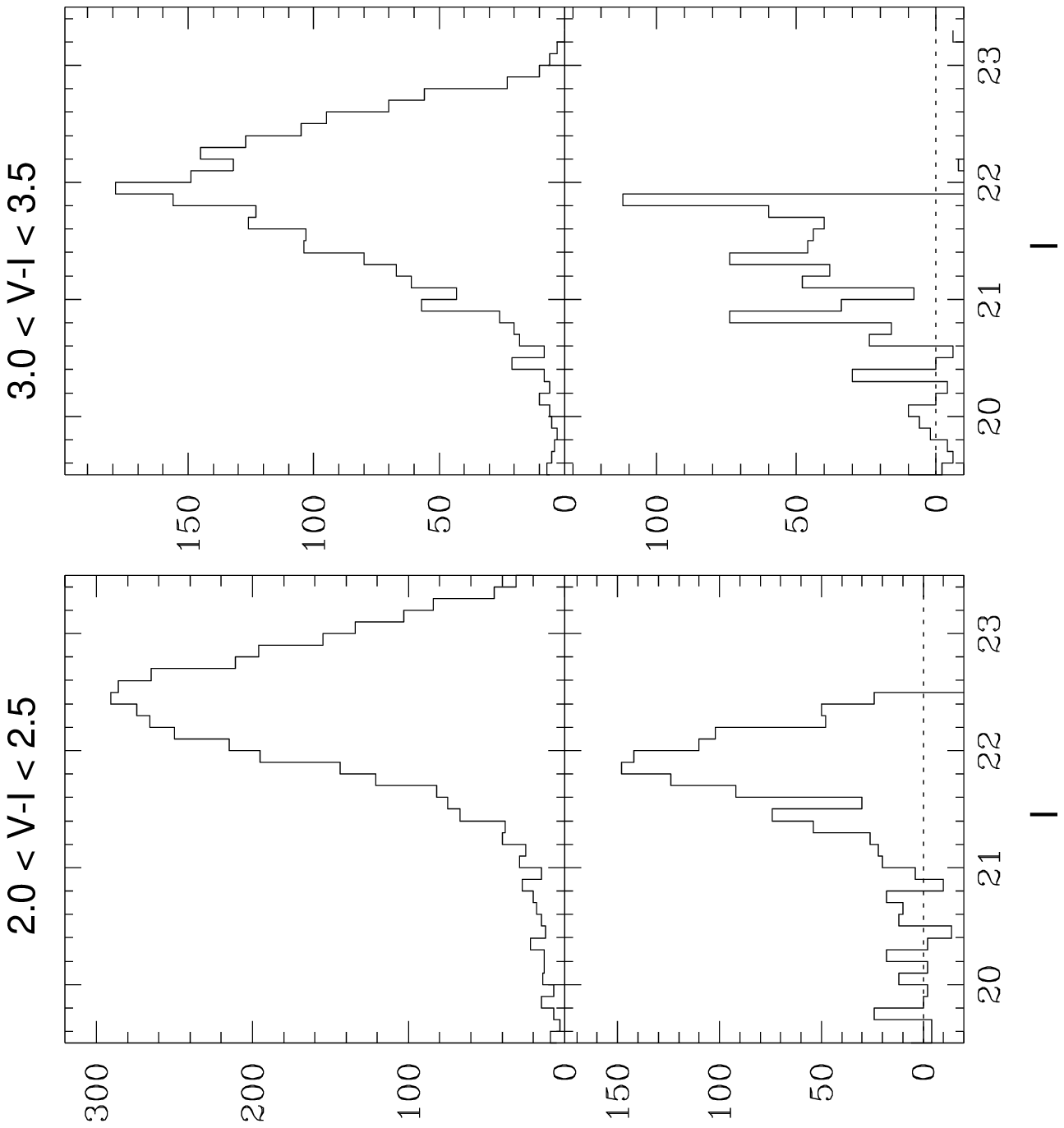}
\end{figure}

\begin{figure}
\plotone{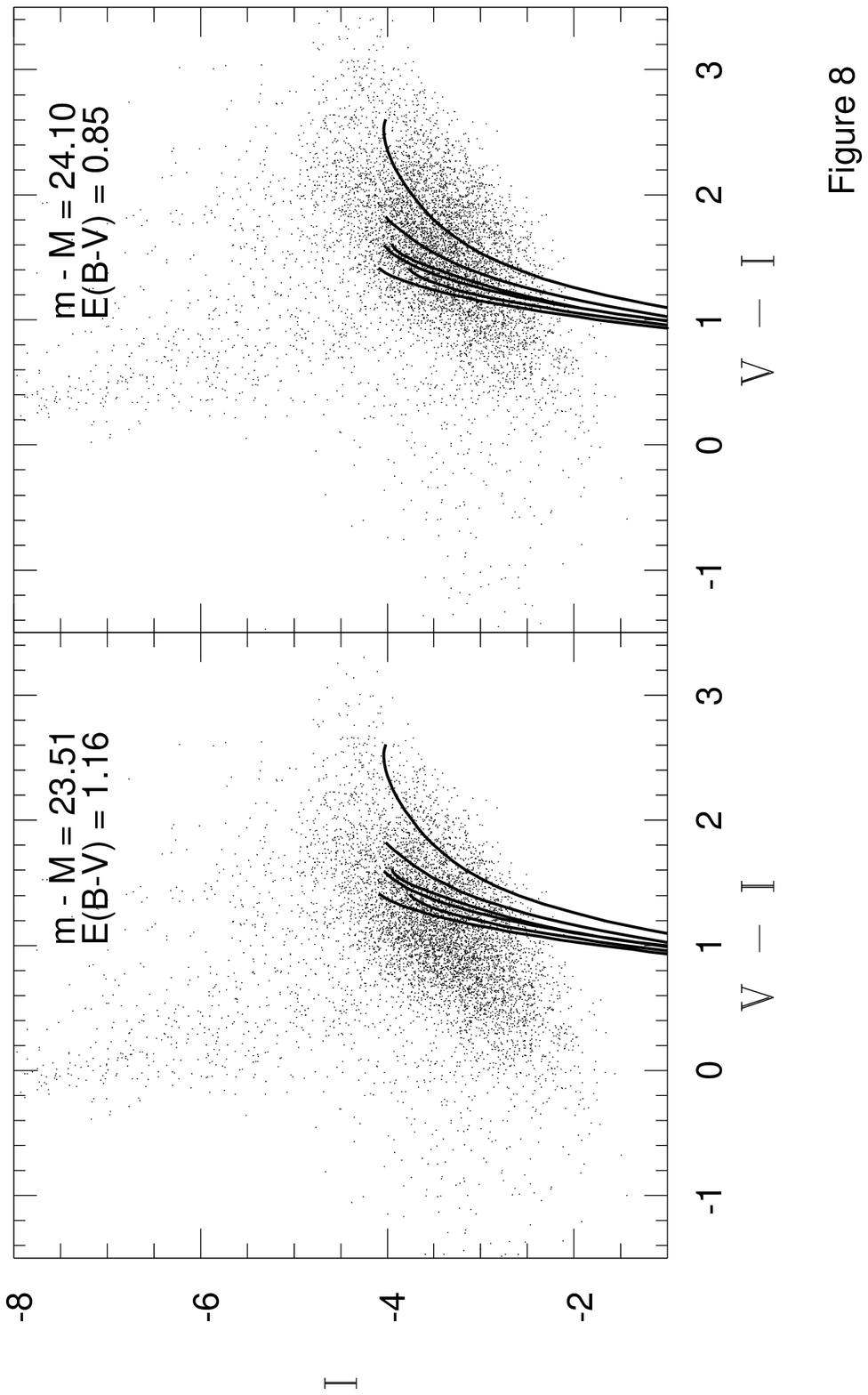}
\end{figure}

\end{document}